\documentclass[%
 reprint,
 amsmath,amssymb,
 aip,jmp,
]{revtex4-2}

\usepackage{graphicx}
\usepackage{dcolumn}
\usepackage{bm}

\begin{document}

\preprint{AIP/123-QED}

\title{Dynamics of the threshold model on hypergraphs}

\author{Xin-Jian Xu}
\affiliation{Department of Mathematics, Shanghai University, Shanghai 200444, China}
\author{Shuang He}
\affiliation{Department of Mathematics, Shanghai University, Shanghai 200444, China}
\author{Li-Jie Zhang}
\email{lijzhang@shu.edu.cn}
\affiliation{Department of Physics, Shanghai University, Shanghai 200444, China}
\date{\today}

\begin{abstract}
The threshold model has been widely adopted as a prototype for studying contagion processes on social networks. In this paper, we consider individual interactions in groups of three or more vertices and study the threshold model on hypergraphs. To understand how high-order interactions affect the breakdown of the system, we develop a theoretical framework based on generating function technology to derive the cascade condition and the giant component of vulnerable vertices, which depend on both hyperedges and hyperdegrees. First, we find a dual role of the hyperedge in propagation: when the average hyperdegree is small, increasing the size of the hyperedges may make the system fragile, while the average hyperdegree is relatively large, the increase of the hyperedges causes the system to be robust. Then, we identify the effects of threshold, hyperdegree, and hyperedge heterogeneities. The heterogeneity of individual thresholds causes the system to be more fragile, while the heterogeneity of individual hyperdegrees or hyperedges increases the robustness of the system. Finally, we show that the higher hyperdegree a vertex has, the larger possibility and faster speed it will get activated. We verify these results by simulating meme spreading on both random hypergraph models and hypergraphs constructed from empirical data.
\end{abstract}


\maketitle


Contagion processes on complex networks occupy a central position in network science~\cite{pastor2015}. A large number of studies have devoted to the literature with focus on pairwise interactions. In the real world, however, interactions can often occur in groups of three or more individuals and cannot be described simply in terms of dyads~\cite{battiston2020}. Motivated by this, we consider the threshold model, one of the typical prototypes for complex contagions, on hypergraphs. In this scenario, a hyperedge can contain certain number of vertices, and the hyperdegree is the number of hyperedges to which it belongs. The state of a vertex can be active or inactive. An inactive vertex will be activated if the active fraction of its neighbors is larger than its adoption threshold. Based on generating function technology, we develop a theoretical framework for the cascade condition and the size of the vulnerable giant component, which are verified by simulations on synthetic and empirical hypergraphs.

\section{Introduction}
Over two decades, complex networks have been used to describe a variety of complex systems. With vertices representing systematic units and edges connected them representing interactions, we obtain real networks spanning many different fields~\cite{torres2018}. Despite differences in their nature, many networked systems may be characterized by similar topological properties. For instance, realistic networks display higher clustering than that expected from a classic random graph. Also, it has been widely observed that the degree distributions of many large networks are heavy tailed. In view of this, Watts and Strogatz~\cite{watts1998} proposed a small-world network model between completely regular networks and completely random networks to describe the properties of high clustering and short distance. On the contrary, Barab\'{a}si and Albert~\cite{barabasi1999} devised a scale-free network model with the power-law degree distribution.

Thanks to these seminal studies, there has been an increasing interest in research in complex networks, among which contagions on networks have attracted much attention from scientists in many fields, such as the spread of infectious diseases~\cite{house2011}, the dissemination of rumors~\cite{zhang2013}, the outbreak of political unrest~\cite{javier2012}, the diffusion of cultural fads~\cite{tan2016}, etc. Yet, individuals in social networks usually exhibit polyadic relationships and cannot be described simply in terms of dyads. Although multiplex networks or simplicial networks have been developed for multimodeling, they are still not enough to provide a complete description of a complex system. Until recently, little attention has been devoted to hypergraphs where the system is described by higher-order interactions.

Bod\'{o} et al.~\cite{bodo2016} first considered communities as hyperedges to analyze a hypergraph-based susceptible-infected-susceptible (SIS) epidemic propagation problem. For an individual belonging to different hyperedges, its infection likelihood is the per contact infection rate multiplied by the sum of the infection pressure brought by different hyperedges, and the infection pressure brought by each hyperedge is a function of the number of infected neighbors of the vertex in this hyperedge. Specifically, they obtained main results based on regular uniform hypergraphs. Another study~\cite{suo2018} exploited RP (spreading to all neighbors) and CP (spreading to the neighbors in a particular hyperedge) strategies in the SIS model on hypergraphs. The generation of the hypergraph follows the idea of the growth mechanism~\cite{zhang2010}, i.e., several newly added vertices and several randomly selected existing vertices are combined to generate hyperedges at each time step. For each vertex, the infection rate is the product of the per contact infection rate and the number of its infected neighbors. On the contrary, de Arruda et al.~\cite{arruda2020} developed a theoretical framework for the SIS model on two peculiar hypergraphs: called hyperblob (a hypergraph composed of a hyperedge containing all vertices in addition to a random regular graph) and hyperstar (a hypergraph composed of a hyperedge containing all vertices in addition to a star graph). As the infection probability is assumed to be independent and identical across successive contacts, the above studies belong to simple contagions. In information propagation, however, both the spreading rate and process are always locally dependent, belonging to complex contagions~\cite{lehmann2018}. In other words, the transmission of social influences or entities often requires contact with multiple sources of activation. In this scenario, recording the relevant lists of interacting vertices gives a more informative picture than reducing these down to a collection of edges, which motivates us to consider hypergraphs.

In this paper, we focus on complex contagions on both uniform and non-uniform hypergraphs. One of the prototypes for studying such dynamics is the threshold model, which can be traced back to 1971 when Schelling~\cite{schelling1971} first considered the problem of residential segregation. Granovetter~\cite{granovetter1978} further developed the model to study collective behavior such as social riots. The name of threshold stems from the step behavior; that is, an individual adopts a new opinion only if more than a certain critical fraction or number of her friends is active. This required fraction/number of adopters in the neighborhood is defined as threshold. Although the propagation rule is simple, the threshold model can exhibit complex behaviors when individual diversity and interaction structure are considered. Watts~\cite{watts2002} first studied the model with one random initiator on complex networks to examine the effects of these factors on the cascade dynamics. It was shown that heterogeneous degrees enhance systemic stability compared to that of homogeneous degrees. Threshold heterogeneity, however, has a contrary effect. Gleeson and Cahalane~\cite{gleeson2007} extended Watts' model to a finite number of initiators and proposed a tree-like approach. Research of the threshold model on multiplex~\cite{brummitt2012} and temporal~\cite{backlund2014} networks has also been carried out; however, there is no general theory explaining how the threshold model on the network of higher-order interactions will be. The goal of this paper is to study the dynamics of the threshold model on general hypergraphs and provide a theoretical explanation.

\section{Threshold model on hypergraphs}
\begin{figure}[t]
\center{\includegraphics[width=0.7\linewidth]{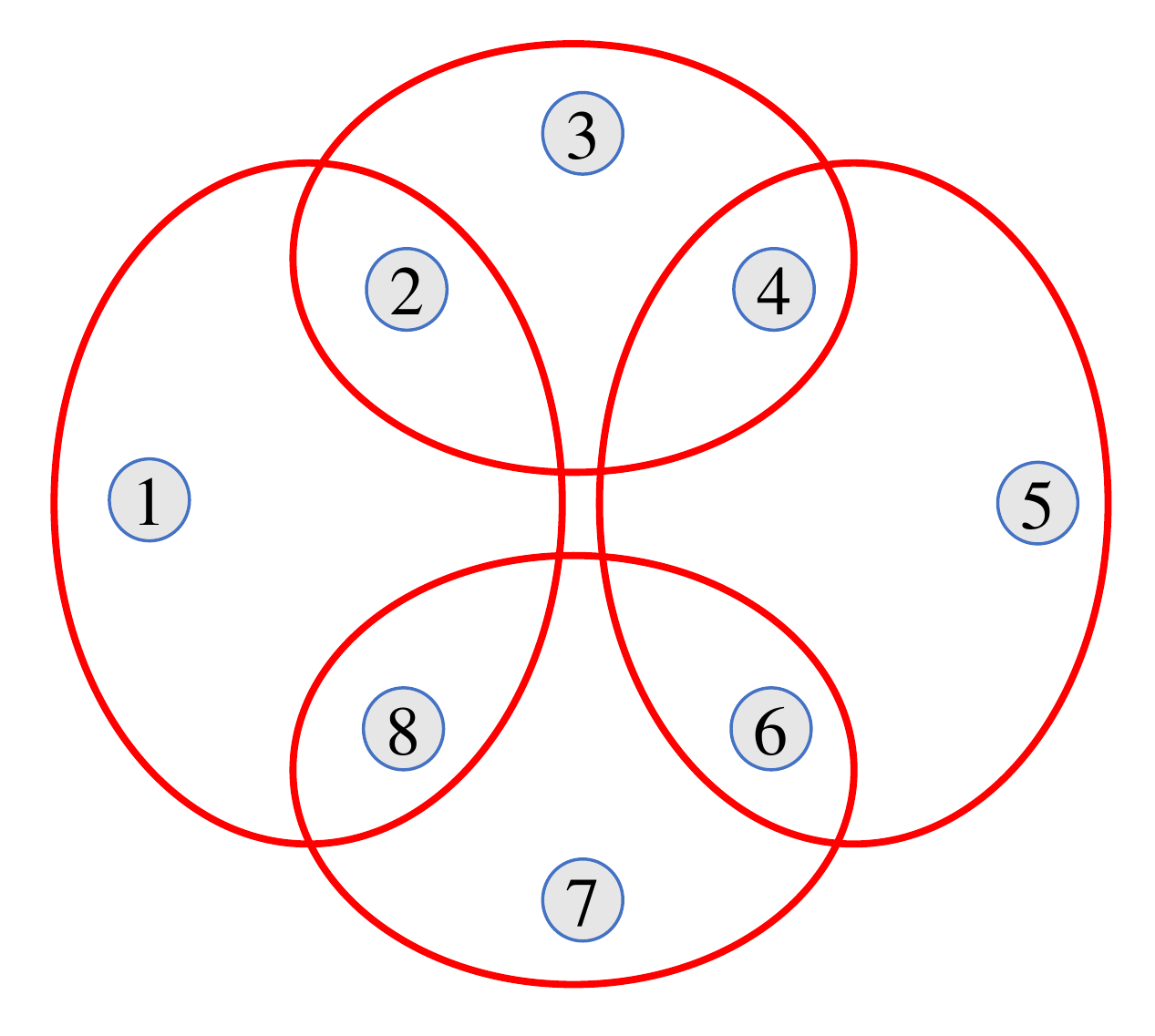}}
\caption{Illustration of a 3-uniform hypergraph with $8$ vertices and $4$ hyperedges.}
\label{hypernet}
\end{figure}

To describe the common properties and characteristic features of hypergraphs, a rigorous and efficient analytic tool is needed, which has been introduced in the form of graph theory in mathematics. A hypergraph consists of vertices connected by hyperedges~\cite{berge1973}. Different from an edge incident to two vertices in a simple graph, a hyperedge can contain arbitrary number of vertices. If there is a vertex set $V=\{v_1,v_2,\cdots,v_N\}$ and a hyperedge set $E=\{e_1,e_2,\cdots,e_M\}$ where each hyperedge is a nonempty subset of the vertex set $V$, i.e., $e_i\subset V,i=1,2,\cdots,M$, then the binary relation $H=(V,E)$ is said to be a hypergraph. The hyperdegree of a vertex is the number of hyperedges to which it belongs; correspondingly, the cardinality of a hyperedge is the number of vertices it contains. If the cardinality of each hyperedge is identical, equal to $d$, then the hypergraph is called $d$-uniform; otherwise, it is non-uniform. In particular, for $d=2$, the hypergraph degenerates into a simple graph. To be more intuitive, a schematic illustration of a uniform hypergraph is shown in Fig.~\ref{hypernet} for $N=8$, $M=4$, and $d=3$. Although the size of hyperedges for any vertex is the same, the hyperdegrees can vary (e.g., see vertices $1$ and $2$).

More generally, one can generate hypergraphs by the configuration model to randomly match the vertices and hyperedges~\cite{bollobas1980}, which is different from the hypergraphs that are regular or generated by the growth model used by previous studies. Due to the property of the configuration model (see Appendix A), the sequences for the hyperdegrees and hyperedges can be arbitrary. Of special interest, we consider Poissonian and power-law distributions of the sequences, respectively.

In the original specification of the threshold model, the dynamics is defined as follows: (i) Initially, each vertex $i$ is assigned a threshold $\mu_i$. One vertex is chosen randomly from the hypergraph to be active, and the others are inactive; and (ii) At each time step, an inactive vertex $i$ will be activated if the active number $m_i$ of its neighbors satisfies
\begin{equation}
\dfrac{m_i}{\kappa_i} \geq \mu_i,
\end{equation}
where $\kappa_i$ is the number of $i$'s neighbors. In a hypergraph, we say that vertex $j$ is a neighbor of vertex $i$ if they belong to at least one common hyperedge. When all vertices have identical threshold, one has $\mu_i=\mu$ and can set $\mu \in [0, 1]$ without loss of generality. While the thresholds vary, they are drawn from a distribution $f(\mu) \in (0, 1)$. In this case, the value of $\mu_i$ can be arbitrary but their mean satisfies $\mu \in [0, 1]$.

Unlike the SIS epidemic model, the vertices in the threshold model can only change from the inactive state to the active state but have no recovery capability thereafter. For a successful cascade, one of the neighbors of the seed must have a threshold $\mu$ no larger than $1/\kappa$. We call such neighbor a vulnerable vertex. For a general hypergraph, the probability of a vertex being vulnerable can be written as
\begin{equation}\label{defvc}
\begin{split}
\rho_\kappa=\left \{
\begin{array}{cc}
    1,& \kappa=0,\\
    P(\mu\leq \frac{1}{\kappa})=\int_0^{\frac{1}{\kappa}} f(\mu){\rm d}\mu, & \kappa>0.
\end{array}
\right.
\end{split}
\end{equation}
Thus, the necessary condition for global cascades is that the component of vulnerable vertices must percolate throughout the whole hypergraph; that is, the largest connected vulnerable cluster must occupy a finite fraction of the hypergraph, which means the presence of the vulnerable giant component.

\section{Generating function approach}
To apply generating function technology, we assume that: (i) the underlying hypergraph is locally tree-like, i.e., it contains no vertex-hyperedge alternative sequences with the same starting and ending vertex; and (ii) the influence of each hyperedge on a vertex is identical.

\subsection{Uniform hypergraphs}

In a $d$-uniform hypergraph, the size of each hyperedge is $d$ and a vertex has hyperdegree $k$ with probability $p_k$, and thus Eq.~\eqref{defvc} reduces to
\begin{equation}
\begin{split}
\rho_k=\left \{
\begin{array}{cc}
    1,& k=0,\\
    P(\mu\leq \frac{1}{k(d-1)})=\int_0^{\frac{1}{k(d-1)}} f(\mu){\rm d}\mu, & k>0.
\end{array}
\right.
\end{split}
\end{equation}
Furthermore, the probability generating function for the hyperdegree distribution of vulnerable vertices is
\begin{equation}\label{fg0}
{G_0(x)=\sum_{k=0}^{\infty}{\rho_k p_k x^k}},
\end{equation}
then one can obtain the fraction of vulnerable vertices $G_0(1)=\sum_{k=0}^{\infty}{\rho_k p_k}$ and the average hyperdegree of vulnerable vertices $G_0^\prime(1)=\sum_{k=1}^{\infty}{k\rho_k p_k}$. To describe propagation from one vertex to another, one also requires the generating function for the excess hyperdegree of vulnerable vertices reached by a random hyperedge, defined by
\begin{equation}\label{fg1}
G_1(x)=\sum_{k=1}^{\infty}{\frac{k\rho_k p_k}{z} x^{k-1}}=\frac{G_0^\prime(x)}{z},
\end{equation}
where $z=\sum_{k=0}^{\infty}{k p_k}$ is the average hyperdegree of all vertices. To analyze the properties of vulnerable clusters, we introduce the generating function for the size distribution of vulnerable components,
\begin{equation}\label{defh0}
H_0(x)=\sum_{n=0}^{\infty}{\alpha_n x^n},
\end{equation}
where $\alpha_n$ is the probability that a randomly chosen vertex belongs to a vulnerable component of size $n$. Analogously, the probability generating function for the size of vulnerable components reached by a random hyperedge
is
\begin{equation}\label{defh1}
H_1(x)=\sum_{n=0}^{\infty}{\beta_n x^n},
\end{equation}
where $\beta_n$ is the probability that a vertex reached by a random hyperedge belongs to a vulnerable component of size $n$.

Let $P(n|k)$ be the probability that when a vertex with hyperdegree $k$ is removed, its $k(d-1)$ neighbors belong to a vulnerable cluster of size summing to exactly $n$, then $P(n-1|k)$ denotes the probability that a vertex with hyperdegree $k$ itself belongs to the vulnerable cluster of size $n$. Therefore, we have
\begin{equation}
\alpha_n=\sum_{k=0}^{\infty}{\rho_k p_k P(n-1|k)}.
\end{equation}
Noting that $\sum_{n=0}^{\infty}{P(n|k) x^n}$ represents the probability generating function of vulnerable clusters to which $k(d-1)$ neighbors belong, by the property of powers of the generating function, it follows that
\begin{equation}
\sum\limits_{n=0}^{\infty}{P(n|k) x^n}=[H_1(x)]^{k(d-1)}.
\end{equation}
Combing these two expressions, $H_0(x)$ can be rewritten as
\begin{equation}
\begin{aligned}
H_0(x)&=\sum_{n=0}^{\infty}{\sum_{k=0}^{\infty}{\rho_k p_k P(n-1|k)}x^n}\\
&=1-G_0(1)+x\sum_{n=0}^{\infty}{\sum_{k=0}^{\infty}{\rho_k p_k P(n|k)} x^n}\\
&=1-G_0(1)+x G_0([H_1 (x)^{d-1}]).
\end{aligned}
\end{equation}
On the other hand, in the limit of a large hypergraph size, if a neighbor vertex of vertex $i$ has hyperdegree $k$, the removal of $i$ actually has no effect on the hyperdegree distribution of the whole hypergraph, which means the probability that the neighbor vertex with hyperdegree $k$ (excess hyperdegree $k-1$) belongs to a vulnerable cluster of size $n$ is $P(n-1|k-1)$. Noting that the hyperdegree $k$ now will obey the excess hyperdegree distribution instead since this neighbor is reached by following a hyperedge from vertex $i$. Therefore, we obtain
\begin{equation}
\beta_n=\sum_{k=1}^{\infty}{\frac{k\rho_k p_k}{z}P(n-1|k-1)}.
\end{equation}
Substituting it into Eq.~\eqref{defh1} yields
\begin{equation}
\begin{split}
H_1(x)&=\sum_{n=0}^{\infty}{\sum_{k=1}^{\infty}{\frac{k\rho_k p_k}{z}P(n-1|k-1)}x^n}\\
&=1-G_1(1)+x\sum_{n=0}^{\infty}{\sum_{k=1}^{\infty}{\frac{k\rho_k p_k}{z}P(n|k-1)}x^n}\\
&=1-G_1(1)+xG_1([H_1(x)]^{d-1}).
\end{split}
\end{equation}

According to the definition of $H_0(x)$, $H_0(1)$ denotes the probability of reaching a finite vulnerable cluster (including the case of reaching a vulnerable cluster of size 0, i.e., reaching a non-vulnerable vertex) from a randomly chosen vertex, so the size of the vulnerable giant component can be written as
\begin{equation}\label{sv}
S_v=1-H_0(1)=G_0(1)-G_0([H_1(1)]^{d-1}).
\end{equation}
Notice that $H_1(1)=1$ is always a trivial solution corresponding to $S_v=0$, we need to find a nontrivial solution between $0$ and $1$ for $H_1(1)$. Let $u=H_1(1)$ and $g(u)=1-G_1(1)+G_1(u^{d-1})$, we rewrite the above equation as
\begin{equation}
u=g(u).
\end{equation}
For the purpose that $y=u$ and $y=g(u)$ intersect between $0$ and $1$, it requires $g^\prime(1)>1$. Finally, the condition for the presence of the vulnerable giant component, i.e., the phase transition point at which the global cascade occurs, can be written as
\begin{equation}
(d-1) G_0^{\prime\prime}(1)=(d-1) \sum_{k=2}^{\infty} k(k-1)p_k \rho_k=\langle k\rangle.\label{condition}
\end{equation}
Particularly, in the case of $d=2$, it degenerates into the cascade condition for dyadic graphs~\cite{watts2002}.

\subsection{Non-uniform hypergraphs}
In a non-uniform hypergraph, both distributions of hyperdegrees and hyperedges should be considered. Assuming that they are characterized by $p_k$ and $q_d$, the corresponding probability generating functions read
\begin{eqnarray}
U_0(x)&=&\sum_{k=0}^{\infty}p_k x^k,\\
V_0(x)&=&\sum_{d=0}^{\infty}q_d x^d.
\end{eqnarray}
To describe propagation from one vertex to another, one also requires probability generating functions for the excess distributions of $U_0(x)$ and $V_0(x)$:
\begin{eqnarray}
U_1(x)&=&\sum_{k=1}^{\infty}\frac{k p_k x^{k-1}}{\langle k\rangle}=\frac{U_0^{\prime}(x)}{\langle k\rangle},\\
V_1(x)&=&\sum_{d=1}^{\infty}\frac{d q_d x^{d-1}}{\langle d\rangle}=\frac{V_0^{\prime}(x)}{\langle d\rangle}.
\end{eqnarray}
Thus, randomly choosing a vertex and the probability generating function for the number of its neighbors is a function composition of $U_0(x)$ and $V_1(x)$,
\begin{equation}\label{ff0}
F_0(x)=U_0(V_1(x))=\sum_{\kappa=0}^{\infty}\eta_\kappa x^\kappa,
\end{equation}
where $\eta_\kappa=\frac{1}{\kappa!}\frac{{\rm d}^\kappa F_0}{{\rm d}x^\kappa}\bigg|_{x=0}$ is the probability that the randomly chosen vertex has $\kappa$ neighbors. Furthermore, random choosing a hyperedge and the probability generating function for the number of its one member's neighbors is
\begin{equation}\label{ff1}
F_1(x)=U_1(V_1(x))=\sum_{\kappa=\langle d\rangle}^{\infty}\lambda_{\kappa-\langle d\rangle} x^{\kappa-\langle d\rangle}.
\end{equation}
Since the expected number of vertices contained in one hyperedge is $\langle d\rangle$, $\lambda_{\kappa-\langle d\rangle}=\frac{1}{(\kappa-\langle d\rangle)!}\frac{{\rm d}^{\kappa-\langle d\rangle} F_1}{{\rm d}x^{\kappa-\langle d\rangle}}\bigg|_{x=0}$ is the probability of other $\kappa-\langle d\rangle$ neighbors of the member.

Next, we consider individual states, and the corresponding probability generating functions are
\begin{eqnarray}
\mathcal{G}_0(x)&=&\sum_{\kappa=0}^{\infty}\rho_{\kappa}\eta_{\kappa}x^{\kappa},\\
\mathcal{G}_1(x)&=&\sum_{\kappa=\langle d\rangle}^{\infty}\rho_{\kappa}\lambda_{\kappa-\langle d\rangle}x^{\kappa-\langle d\rangle}.
\end{eqnarray}
Furthermore, denoting with $\mathcal{H}_0(x)$ and $\mathcal{H}_1(x)$ the probability generating functions for the vulnerable components reached, respectively, by one vertex and one hyperedge, we obtain the self-consistent equations,
\begin{eqnarray}
\mathcal{H}_0(x)&=&1-\mathcal{G}_0(1)+x\mathcal{G}_0(\mathcal{H}_1(x)),\\
\mathcal{H}_1(x)&=&1-\mathcal{G}_1(1)+x\mathcal{G}_1(\mathcal{H}_1(x)).
\end{eqnarray}
Therefore, the cascade condition can be rewritten as
\begin{equation}\label{nonucd}
\mathcal{G}_1^{\prime}(1)=\sum_{\kappa=\langle d\rangle}^{\infty}\rho_\kappa(\kappa-\langle d\rangle)\lambda_{\kappa-\langle d\rangle}=1,
\end{equation}
and the corresponding size of the vulnerable component giant reads
\begin{equation}\label{nonusv}
\mathcal{S}_v=1-\mathcal{H}_0(1)=\mathcal{G}_0(1)-\mathcal{G}_0(\mathcal{H}_1(1)).
\end{equation}

In what follows, we provide two examples. First, we assume that both distributions of hyperdegrees and hyperedges are Poissonian; i.e., $p_k=e^{-\langle k\rangle}\langle k\rangle^k/k!$ and $q_d=e^{-\langle d\rangle}\langle d\rangle^d/d!$. It is easy to obtain $U_0(x)=U_1(x)=e^{\langle k\rangle(x-1)}$, $V_0(x)=V_1(x)=e^{\langle d\rangle(x-1)}$, and $F_0(x)=F_1(x)=e^{\langle k\rangle[e^{\langle d\rangle(x-1)}-1]}$. According to Eqs.~\eqref{ff0} and~\eqref{ff1}, it follows that
\begin{eqnarray}
\eta_{\kappa}&=&e^{(\langle k\rangle e^{-\langle d\rangle}-\langle k\rangle)}\theta_{\kappa}\frac{\langle d\rangle^{\kappa}}{{\kappa}!},\\
\lambda_{\kappa-\langle d\rangle}&=&\eta_{\kappa-\langle d\rangle},
\end{eqnarray}
with $\theta_{\kappa}=\sum_{\nu=1}^{\kappa}\sum_{\pi=0}^{\nu}\frac{1}{\nu!}{(-1)}^{\pi} \binom{\nu}{\pi}{(\nu-\pi)}^{\kappa}{(\langle k\rangle e^{-\langle d\rangle})}^\nu$. Substituting the above expressions into Eqs.~\eqref{nonucd} and~\eqref{nonusv} yields the condition for a global cascade and the corresponding size. Second, we replace the hyperedge distribution by a power law, i.e., $q_d=d^{-\gamma}/\zeta(\gamma)$ with the exponent $\gamma$ and the normalization $\zeta(\gamma)$, and have
\begin{equation}
\langle d\rangle=\sum_{d=1}^{\infty}\frac{d^{-(\gamma-1)}}{\zeta(\gamma)}=\frac{{\rm Li}_{\gamma-1}(1)}{{\rm Li}_{\gamma}(1)},
\end{equation}
where ${\rm Li}_{\gamma}(x)=\sum_{d=1}^{\infty}d^{-\gamma}x^{d}$ is the polylogarithm function. It is easy to obtain
\begin{eqnarray}
V_0(x)&=&\frac{{\rm Li}_{\gamma}(x)}{{\rm Li}_{\gamma}(1)},\\
V_1(x)&=&\frac{V_0^{\prime}(x)}{\langle d\rangle}=\frac{{\rm Li}_{\gamma}^{\prime}(x)}{{\rm Li}_{\gamma-1}(1)},\\
U_0(x)&=&U_1(x)=e^{\langle k\rangle(x-1)},\\
F_0(x)&=&F_1(x)=e^{\langle k\rangle\left[\frac{{\rm Li}_{\gamma}^{\prime}(x)}{{\rm Li}_{\gamma-1}(1)}-1\right]},
\end{eqnarray}
which leads to
\begin{eqnarray}
\eta_{\kappa}&=&\frac{1}{\kappa!}\frac{{\rm d}^{\kappa}}{{\rm d} x^{\kappa}}\left[e^{\langle k\rangle (\frac{{\rm Li}_{\gamma}^{\prime}(x)}{{\rm Li}_{\gamma-1}(1)}-1)}\right]\bigg|_{x=0},\\
\lambda_{\kappa-\langle d\rangle}&=&\frac{1}{(\kappa-\langle d\rangle)!}\frac{{\rm d}^{\kappa-\langle d\rangle}}{{\rm d} x^{\kappa-\langle d\rangle}}\left[e^{\langle k\rangle(\frac{{\rm Li}_{\gamma}^{\prime}(x)}{{\rm Li}_{\gamma-1}(1)}-1)}\right]\bigg|_{x=0}.
\end{eqnarray}

\section{Results of synthetic hypergraphs}

\begin{figure}[t]
\center{\includegraphics[height=0.5\textwidth]{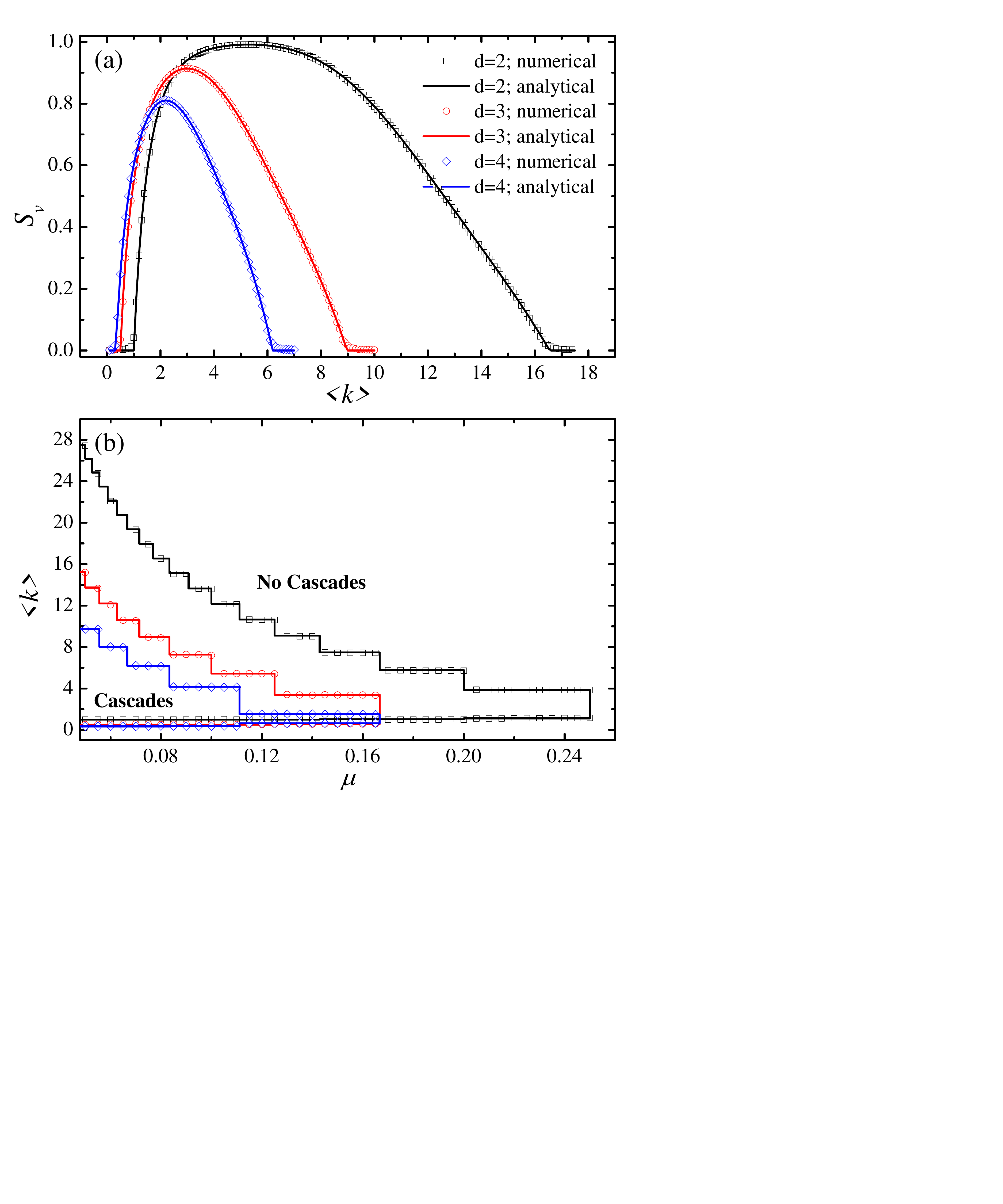}}
\vspace{-0.5cm}
\caption{Vulnerable giant components $S_v$ as a function of the average connectivity $\langle k\rangle$ in Poissonian uniform hypergraphs for $\mu=0.08$ (a) and cascade windows in the ($\mu, \langle k\rangle$) plane inside which the breakdown of the system is observed (b). $100$ simulations were performed on three Poissonian uniform hypergraphs ($N = 10000$) with hyperedge sizes of $d=2$, $3$, and $4$, respectively.}\label{hercw}
\end{figure}

\begin{figure}[t]
\center{\includegraphics[height=0.75\textwidth]{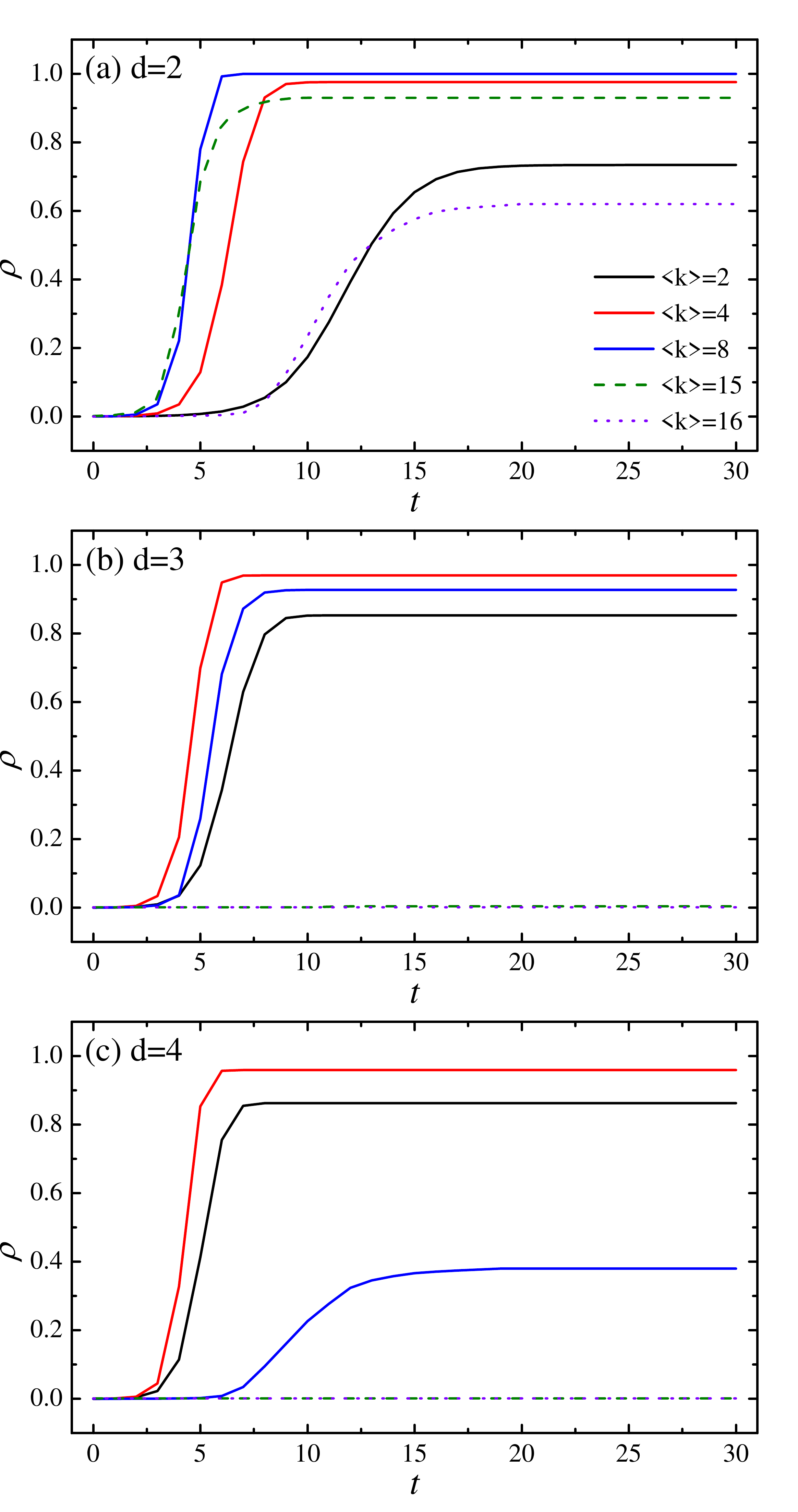}}
\vspace{-0.5cm}
\caption{Temporal evolution of the active fraction in Poissonian uniform hypergraphs for $d=2$ (a), $3$ (b), and $4$ (c), respectively.}\label{herfz}
\end{figure}

\begin{figure}[t]
\center{\includegraphics[height=0.75\textwidth]{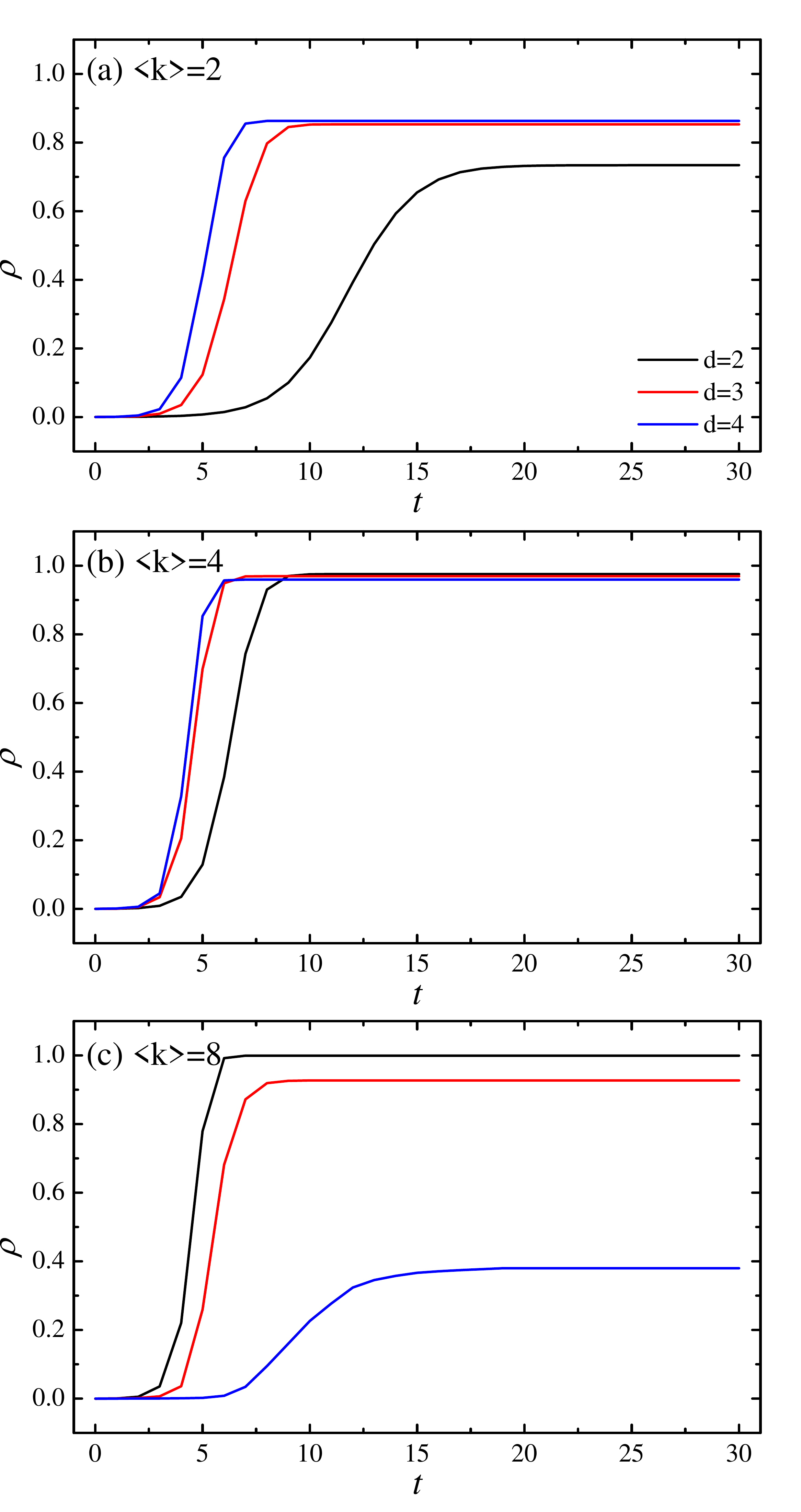}}
\vspace{-0.5cm}
\caption{Temporal evolution of the active fraction in Poissonian uniform hypergraphs for $\langle k\rangle=2$ (a), $4$ (b), and $8$ (c), respectively.}\label{herfd}
\end{figure}

\subsection{Uniform hypergraphs}
We first specify that all vertices in a hypergraph have identical threshold $\mu$. For a vertex $i$ of hyperdegree $k_i$, whether it is active or not depends on the value of $k_i(d-1)$. Therefore, for the whole hypergraph, whether the vulnerable giant component $S_v$ appears or not is closely related to the average hyperdegree $\langle k\rangle$ and the hyperedge size $d$.

Figure~\ref{hercw}(a) plots $S_v$ as a function of $\langle k\rangle$ in uniform hypergraphs under the personal threshold $\mu=0.08$. The hyperdegree distribution is Poissonian. Open points represent simulation results and solid lines are theoretical predictions of Eq.~\eqref{sv} (see Appendix B for details). Apparently, a good agreement is observed. When $\langle k\rangle$ is small (low connectivity regime), the hypergraph is very sparse and propagation is not possible. As $\langle k\rangle$ increases, $S_v$ changes from $0$ to greater than $0$ and the global cascade starts to occur in the hypergraph, corresponding to a critical value of the average hyperdegree called the lower boundary of the cascade. Naturally, larger $z$ can increase the connectivity among vulnerable vertices, leading to a larger $S_v$. However, when $\langle k\rangle$ is very large (high connectivity regime), the vulnerable condition is so difficult to satisfy; i.e., the number of vulnerable vertices will decrease and the global cascade will no longer occur, corresponding to a critical value of the average hyperdegree called the upper boundary. Thus, only within an intermediate range of $\langle k\rangle$ can a global cascade be triggered given a proper value $\mu$. As demonstrated in Fig.~\ref{hercw}(b), the cascade condition (Eq.~\eqref{condition}) is expressed as a boundary (solid lines) in the ($\mu, \langle k\rangle$) plane. To obtain numerical results, we first obtain two transition points of $\langle k\rangle$ in Fig.~\ref{hercw}(a) for the given $\mu$. Then, we change the value of $\mu$ and plot the boundary in Fig.~\ref{hercw}(b).

The results shown in Figs.~\ref{herfz} and Fig.~\ref{herfd} support such phenomenon in a more detailed manner. Still for the $d$-uniform hypergraph with a Poissonian hyperdegree distribution, randomly given one initial seed, we plot time evolution of the fraction of active vertices with different hyperdegrees for each value of $d$ in Fig.~\ref{herfz}. As $\langle k\rangle$ increases beyond the lower boundary, the spread is accelerated at first and the corresponding active fraction at steady state increases, which is caused by the increased connectivity of the hypergraph. However, when $\langle k\rangle$ approaches or exceeds the upper boundary, the spreading slows down instead and the corresponding active fraction at steady state decreases too, which is due to the gradual decrease of vulnerable vertices. The effect of $d$ on cascade behavior is illustrated in Fig.~\ref{herfd}. We also observe dual behavior as in Fig.~\ref{herfz}. When $\langle k\rangle$ is small ($\langle k\rangle=2$), the vulnerable condition is easy to be satisfied for all values of $d$. Although the increase of $d$ causes a cascade to be more likely to happen, the lower connectivity of the hypergraph weakens this effect. While $\langle k\rangle$ is relatively large ($\langle k\rangle=8$), the dominance of local stability of a vertex makes it difficult to be vulnerable, and the increase of $d$ further reduces the likelihood of activation.

\begin{figure}[t]
\center{\includegraphics[height=0.75\textwidth]{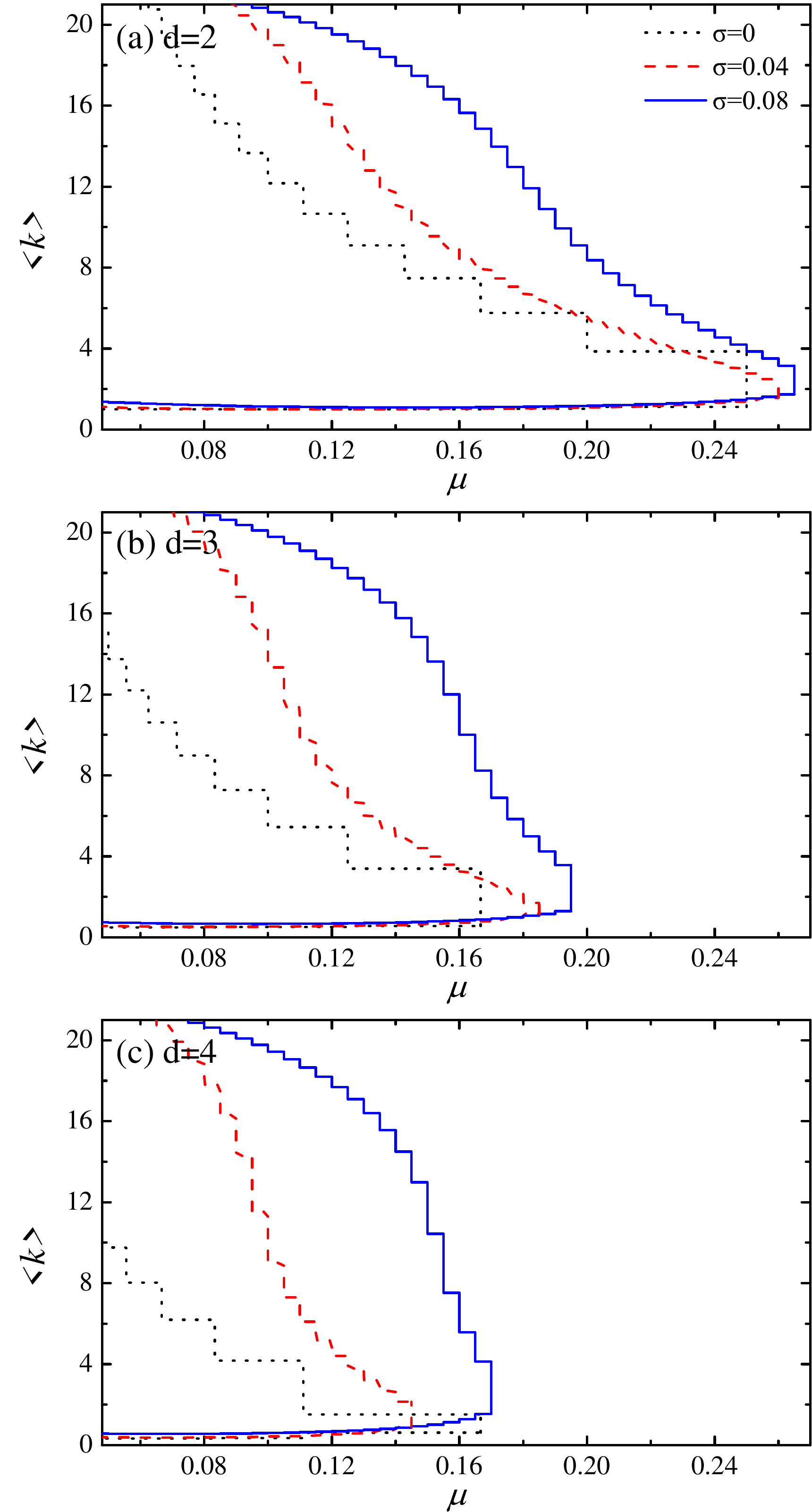}}
\vspace{-0.5cm}
\caption{Effect of heterogeneous thresholds on cascade windows in Poissonian uniform hypergraphs for $d=2$ (a), $3$ (b), and $4$ (c), respectively.}\label{hett}
\end{figure}

\begin{figure}[t]
\center{\includegraphics[height=0.75\textwidth]{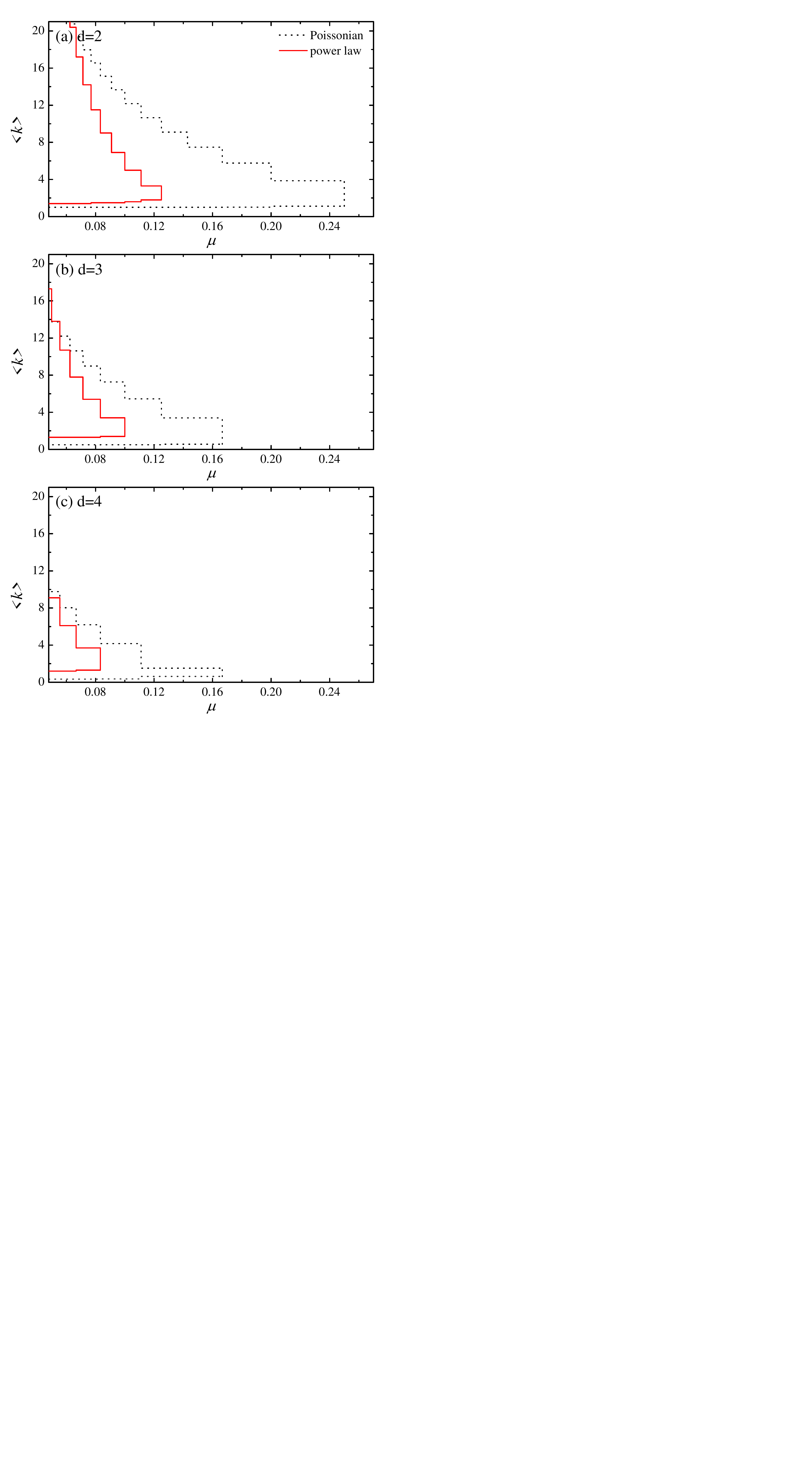}}
\vspace{-0.5cm}
\caption{Effect of heterogeneous hyperdegrees on cascade windows in power-law uniform hypergraphs for $d=2$ (a), $3$ (b), and $4$ (c), respectively.}\label{hetd}
\end{figure}

The above results are obtained based on the identical threshold and the Poissonian distribution. In reality, however,
each person has different extent of accepting a new entity. What effect will be caused by such threshold heterogeneity? Here, we address this issue by considering that the threshold obeys the Gaussian distribution with mean $\mu$ and standard deviation $\sigma$. When $\sigma=0$, it reduces to the case that all vertices' thresholds are the same. Figure~\ref{hett} shows the cascade windows for $\sigma=0.04$ and $0.08$, respectively. Regardless of the hyperedge size, the cascade window becomes larger as $\sigma$ increases, which indicates that the heterogeneity of thresholds leads to a less stable system. On the other hand, in real life, individuals may have different numbers of friends or various communities which they belong to, which means that hyperdegrees may vary from one to another. Therefore, the Poissonian distribution that generates similar hyperdegrees may not be enough to describe such a phenomenon and the heterogeneity of hyperdegrees should be taken into consideration. A power-law distribution is such a distribution that can represent this heterogeneity. As demonstrated in Fig.~\ref{hetd}, the cascade windows under power-law distributions is always smaller than those under Poissonian distributions in spite of the hyperedge size, which implies that heterogeneous hyperdegrees increase the robustness of the system.

\subsection{Non-uniform hypergraphs}

\begin{figure}[t]
\center{\includegraphics[height=0.5\textwidth]{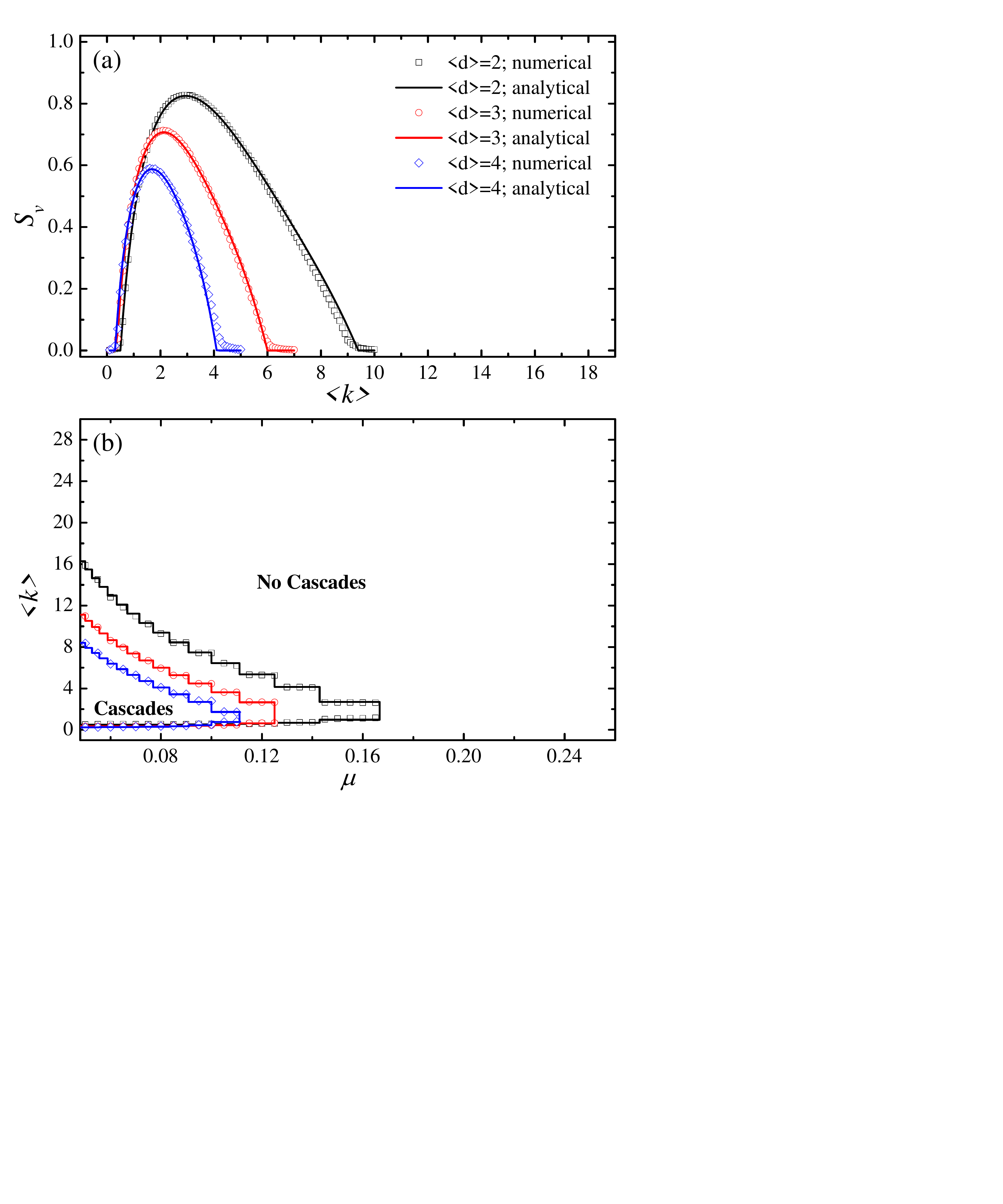}}
\vspace{-0.5cm}
\caption{Vulnerable giant components $S_v$ as a function of the average connectivity $\langle k\rangle$ in non-uniform hypergraphs for $\mu=0.08$ (a) and cascade windows in the ($\mu, \langle k\rangle$) plane inside which the breakdown of the system is observed (b). Both distributions of hyperdegrees and hyperedges are Poissonian. $100$ simulations were performed on the hypergraphs ($N = 10000$) with $\langle d\rangle=2$, $3$, and $4$, respectively.}\label{herncw}
\end{figure}

We begin with the simple case that both hyperdegrees and hyperedges are Poissonian. Figure~\ref{herncw}(a) shows $S_v$ as a function of $\langle k\rangle$. Under the identical threshold $\mu=0.08$, we notice a non-monotonic relationship between them. That is, as $\langle k\rangle$ increases, $S_v$ becomes larger from zero at first and gradually decreases to zero finally. Similar to the uniform hypergraph, the difference in the value of $\langle d\rangle$ of the non-uniform hypergraph would lead to an apparent variation of the transition point on the upper boundary, while it causes a tiny change on the lower boundary. Figure~\ref{herncw}(b) demonstrates the cascade window for different average sizes $\langle d\rangle$ of hyperedges. Again, we notice the dual effect of $\langle d\rangle$: for a small value of $\langle k\rangle$, the increase of $\langle d\rangle$ causes the system to be a little vulnerable, whereas for a large value of $\langle k\rangle$, the increase of $\langle d\rangle$ would make the system more robust.

\begin{figure}[h]
\center{\includegraphics[height=0.75\textwidth]{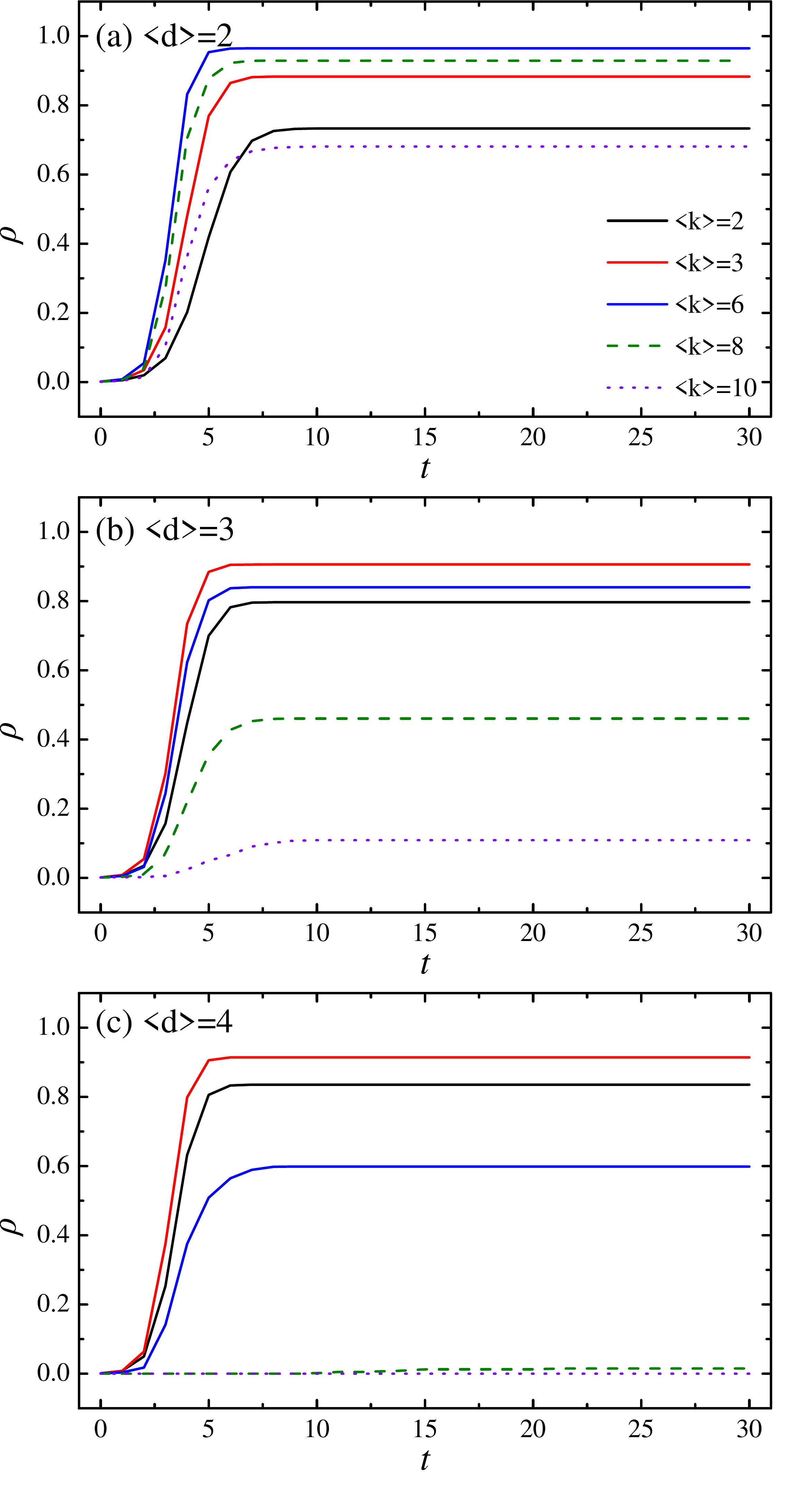}}
\vspace{-0.5cm}
\caption{Temporal evolution of the active fraction in non-uniform hypergraphs for $\langle d\rangle=2$ (a), $3$ (b), and $4$ (c), respectively. Both distributions of hyperdegrees and hyperedges are Poissonian.}\label{hernfz}
\end{figure}

\begin{figure}[h]
\center{\includegraphics[height=0.75\textwidth]{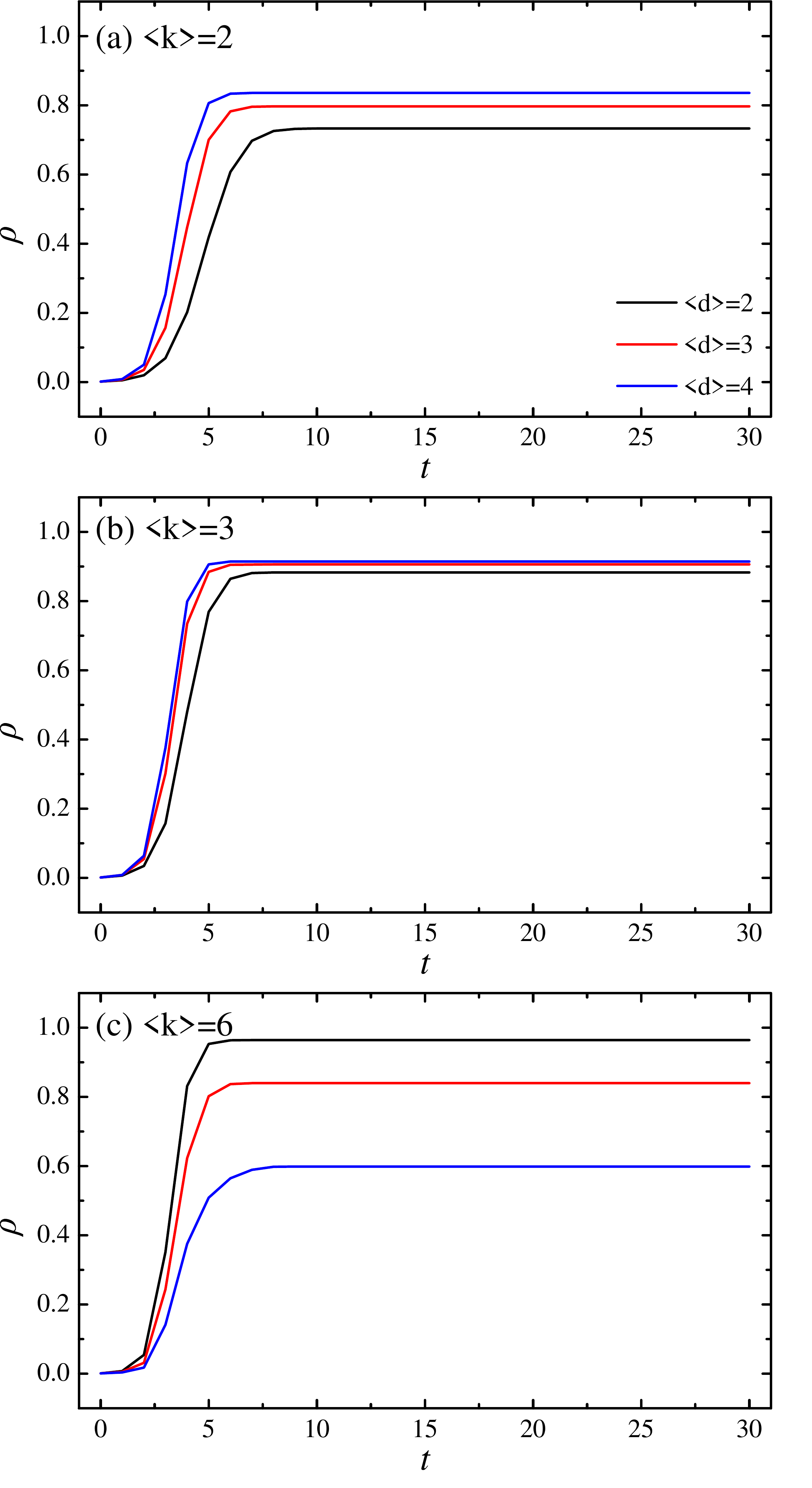}}
\vspace{-0.5cm}
\caption{Temporal evolution of the active fraction in non-uniform hypergraphs for $\langle k\rangle=2$ (a), $3$ (b), and $6$ (c), respectively. Both distributions of hyperdegrees and hyperedges are Poissonian.}\label{hernfd}
\end{figure}

To get a deeper insight into the dynamic behavior, we simulate the temporal evolution of active vertices under $\mu=0.08$. As shown in Fig.~\ref{hernfz}, given the average size of hyperedges, a larger value of $\langle k\rangle$ usually results in a larger active fraction and a faster contagion speed at first. However, once $\langle k\rangle$ is close to or beyond the upper boundary, the increase of $\langle k\rangle$ would decrease $\rho$ as well as the speed. On the other hand, given the average hyperdegree of vertices, the non-monotonic effect of $\langle d\rangle$ is demonstrated in Fig.~\ref{hernfd}. When the average hyperdegree is small ($\langle k\rangle$=2), the increase of $\langle d\rangle$ enhances the connectivity of the hypergraph, which gives rise to contagion. While for the large average hyperdegree ($\langle k\rangle$=6), the hypergraph is relatively dense, and the increase of $\langle d\rangle$ results in vertices being more difficult to be vulnerable and, therefore, the reduction of the active fraction.

\begin{figure}[h]
\center{\includegraphics[height=0.75\textwidth]{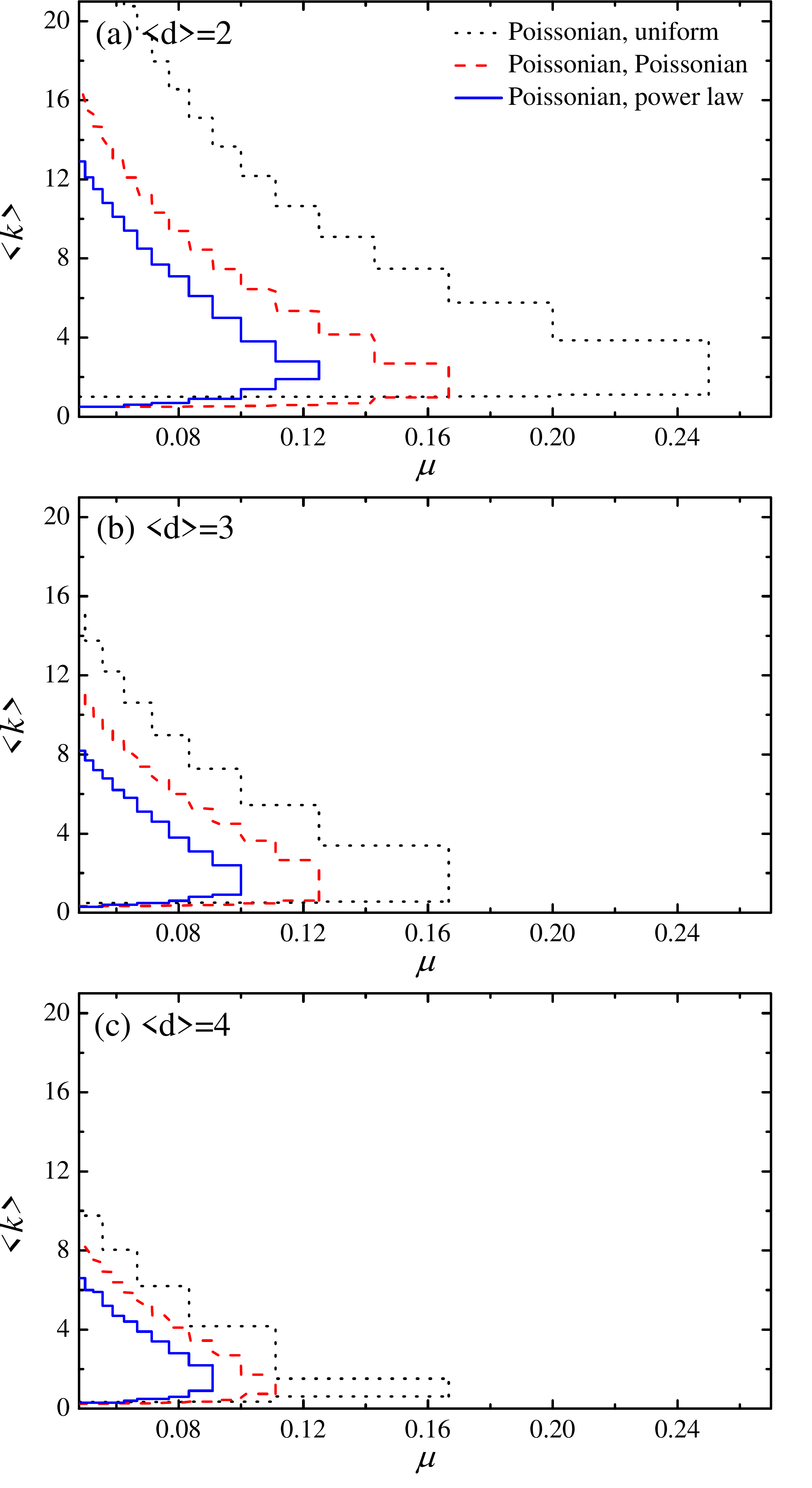}}
\vspace{-0.5cm}
\caption{Effect of heterogeneous hyperedges on cascade windows in non-uniform hypergraphs for $\langle d\rangle=2$ (a), $3$ (b), and $4$ (c), respectively.}\label{hetns}
\end{figure}

Next, we consider the non-uniform hypergraphs with power-law distributions of hyperedges while the Poissonian distribution of hyperdegrees is preserved. For comparison, the results of the uniform and Poissonian distributions of the hyperedges are also presented. The cascade windows are shown in Fig.~\ref{hetns}. Obviously, the more heterogeneous the hyperedges are, the more robust the system is against contagions.

\begin{figure}[t]
\center{\includegraphics[height=0.75\textwidth]{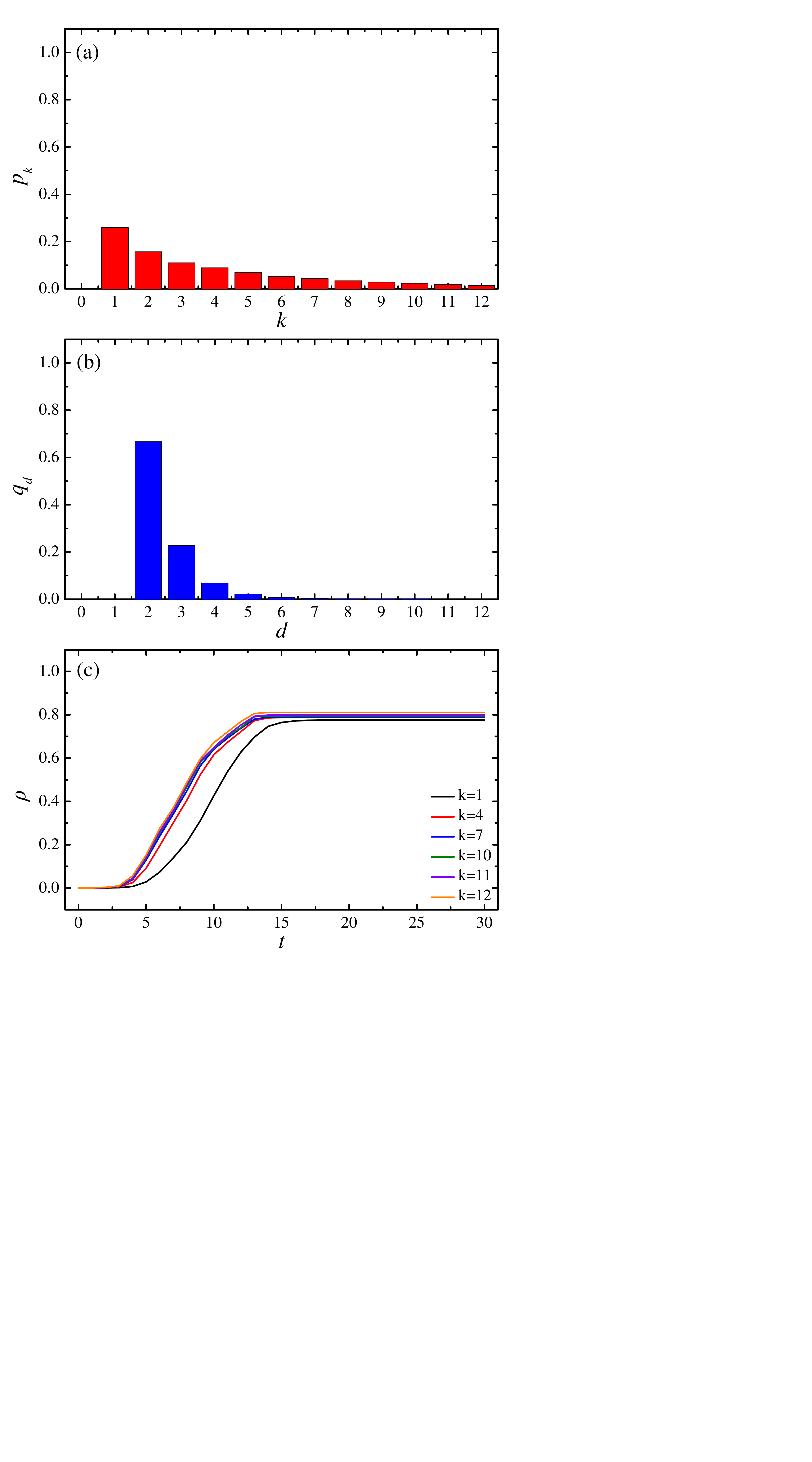}}
\caption{Distributions of hyperdegrees (a) and hyperedges (b) of the hypergraph generated by the Deezer Europe Social Relationships and the temporal dynamics of active vertices (c).}\label{deezer}
\end{figure}

\begin{figure}[t]
\center{\includegraphics[height=0.75\textwidth]{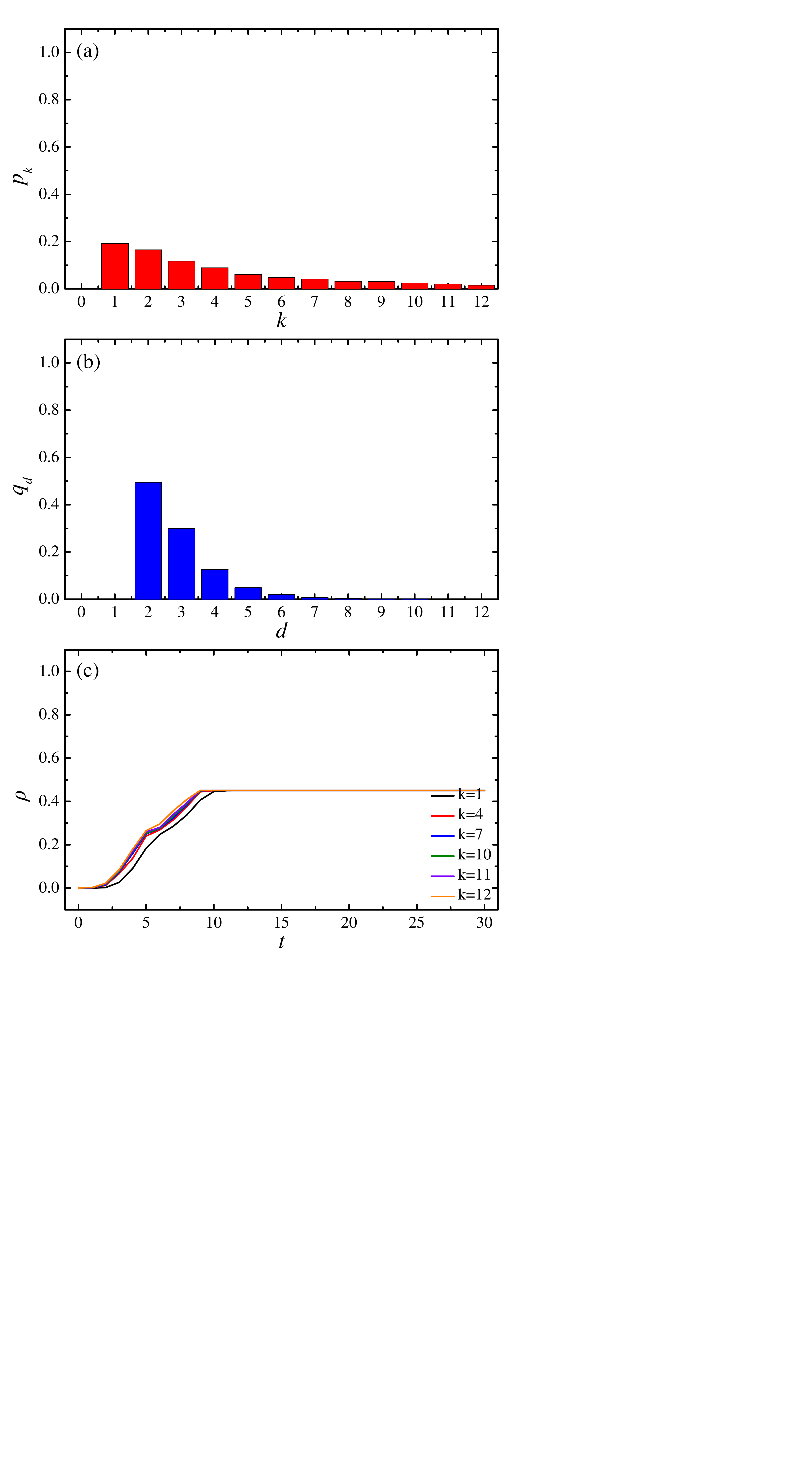}}
\caption{Distributions of hyperdegrees (a) and hyperedges (b) of the hypergraph generated by the Twitch user-user relationships and the temporal dynamics of active vertices (c).}\label{twitch}
\end{figure}

\section{Results of empirical hypergraphs}

We start by examining the threshold model on the hypergraph of Deezer Europe Social Relationships from an online music website. The data, available at https://snap.stanford.edu/data/, were collected from the public API in March 2020. Vertices are Deezer users from European countries and edges are mutual follower relationships between them. Then, a dyadic graph is obtained with $28281$ vertices and $92752$ edges. To represent polyadic relations, we adopt the concept of clique~\cite{bodo2016,palla2005}. If $k$ users share one moment, then a completely connected subgraph on $k$ vertices is generated. From the dyadic graph, we find all complete subgraphs and regard them as hyperedges, and finally obtain the hypergraph. In this way, the generated hypergraph has $66155$ hyperedges, including $44099$ 2-cliques, $15013$ 3-cliques, $4592$ 4-cliques, $1498$ 5-cliques, $543$ 6-cliques, $273$ 7-cliques, $84$ 8-cliques, and very few cliques beyond $8$. Based on this, we obtain the average hyperdegree $\langle k\rangle=5.8320$ and the average hyperedge size $\langle d\rangle=2.5008$

As shown in Figs.~\ref{deezer}(a) and ~\ref{deezer}(b), both distributions of the hyperdegree and hyperedge size are closer to the power law than the Poissonian, displaying the non-uniform characteristics of empirical hypergraphs. Because of the heterogeneous hyperdegrees, we compute the fraction of active vertices of hyperdegree $k$ and present the results in Fig.~\ref{deezer}(c). As one can see, the larger the hyperdegree of $k$, the higher level it reaches with less time. We argue that the main reason is the emergence of a small number of hubs that play important roles in contagions. On the contrary, most vertices possess lower hyperdegrees. Due to the locally tree-like structure, even though these vertices are vulnerable, it is less likely for them to activate other vertices since they belong to fewer hyperedges, which results in the lower active fraction at each time step.

We now examine the threshold model on another empirical hypergraph of Twitch user-user relationships of gamers who stream in a certain language (we choose English for the case study). The data are also available at https://snap.stanford.edu/data/ and were collected in May 2018. In this dataset, vertices are the users themselves and the edges are mutual friendships between them. Then, a dyadic graph is generated with $7126$ vertices and $35324$ edges. Using the same method, we finally obtain $26252$ hyperedges, mainly including $13004$ 2-cliques, $7855$ 3-cliques, $3301$ 4-cliques, $1271$ 5-cliques, $511$ 6-cliques, $168$ 7-cliques, $99$ 8-cliques, $29$ 9-cliques, and very few cliques beyond $9$. The average hyperdegree is $\langle k\rangle=10.4444$ and the average size of the hyperedges is $\langle d\rangle=2.8405$.

Figures~\ref{twitch}(a) and~\ref{twitch}(b) show the distributions of hyperdegrees and hyperedge sizes, which also approach to the power law. Therefore, the active fraction of higher-hyperdegree vertices always grows faster (see Fig.~\ref{twitch}(c)). However, the difference among the fractions for different values of $k$ is tiny, which is mainly because the heterogeneity in this empirical hypergraph is weaker than the previous one, as one can see from two distributions of $p_k$ and $q_d$.

\section{Conclusion}
Research in network science has involved with many fields of complex systems, among which social contagions keep rising. However, existing studies mainly consider pairwise interactions between agents, which violates real social networks that interpersonal interactions are often higher-order. It is, therefore, of great importance to develop new models and methods to explain such behavior in a quantitative way.

In this paper, we have extended the classic threshold model to hypergraphs. Based on the assumptions of a locally tree-like structure and identical influence of hyperedges, we developed a generating function method to calculate the vulnerable giant component and provided numerical simulations for arbitrary uniform and non-uniform hypergraphs. We obtained the condition and prevalence of large cascades in the hypergraph. On one hand, we observed the same behavior as that in dyadic graphs: the heterogeneity of individual thresholds leads to a more fragile system, while the heterogeneity of individual hyperdegrees increases the robustness of the system. On the other hand, we observed unique behavior for hypergraphs. When the average hyperdegree $\langle k\rangle$ is very small, the systematic robustness reduces slightly as $\langle d\rangle$ increases. While it is relatively large, the increase of $\langle d\rangle$ prevents global cascades more significantly. Moreover, the heterogeneity of hyperedges in the non-uniform hypergraph enhances the systematic robustness in comparison with the uniform hypergraph. Finally, we simulated the threshold model on two empirical hypergraphs. We find that vertices with larger hyperdegree $k$ speed up the spreading process, and the corresponding active fraction reaches higher level at the steady state. Our results complement the literature of contagion processes on complex networks, although they are derived from locally tree-like hypergraphs, different from hypernetworks~\cite{sorren2012}.

In the future, it will be interesting to consider more contagion mechanisms to make the present model more feasible for the real world. First, memory of non-redundant information plays an important role in reinforcement~\cite{wang2015}. It is interesting to incorporate the correlation between local structures and social reinforcement. Second, social interactions are usually asymmetric; namely, the transmission depends on both the probability of giving it and the probability of catching it~\cite{huang2016}. A natural approach is to adopt the double-threshold model. Finally, the standard threshold model usually focuses on the final consequence of a contagion. To get more information of dynamic properties, one can build the continuous-time stochastic version of the model~\cite{ran2020}.

\section*{Appendix A: Simulation of the systematic dynamics}

Simulation procedure includes two parts: the generation of the underlying hypergraph and the evolution of the threshold model.

We generate the underlying hypergraph by the configuration model~\cite{bollobas1980}: (i) A priori sequence of random integers $k_i$, each of which represents the hyperdegree of a vertex, is drawn from a distribution function $p_k$; (ii) For the uniform hypergraph, specifying the hyperedge size $d$, one can obtain $\sum_i k_i/d$ hyperedges. If it is not integral, keep increasing the hyperdegree of the first vertex. Then, a hyperedge sequence is obtained, each of which would contain $d$ vertices. For the non-uniform case, generate a sequence of random integers $d_j$ from a distribution function $q_d$, each of which represents the size of a hyperedge. The two sequences satisfy $\sum_i k_i=\sum_j d_j$; and (iii) Randomly match the vertices and hyperedges from the two sequences, and finally a hypergraph is generated.

We simulate the threshold model as follows: i) Initially randomly choose a vertex to be active, while the others are inactive. For each vertex $i$, a threshold $\mu_i$ is drawn from $f(\mu)$ to represents its rationality; ii) At each time step, each inactive vertex will check its neighbors' states and count $m_i$ according to the number of active neighbors. If the activating condition (Eq. (1)) is satisfied, it becomes active. Here, parallel updating is utilized. Compute the fraction of active vertices; and iii) Repeat ii) until no further activation occurs.

\section*{Appendix B: Calculation of the vulnerable giant component}

Here, we just take the uniform hypergraph for example. Given the threshold $\mu$, the hyperedge size $d$, the average hyperdegree $\langle k\rangle$, and the distribution $p_k$, one can calculate $S_v$ theoretically based on Eq.~\eqref{sv}. To compute $S_v$ numerically, we first set $m_i=1$ $(i=1,\cdots,N)$ and determine whether each vertex is vulnerable given $\mu$. Then, delete all non-vulnerable vertices and corresponding hyperedges connected to them, yielding a new hypergraph. Finally, apply the breath first search algorithm to the generated hypergraph to find the giant component, the size of which is exactly $S_v$.

\section*{Acknowledgments}
This work was supported by the Natural Science Foundation of China under Grant Nos. 12071281 and 11771277.




\begin{references}

\bibitem{pastor2015}
R. Pastor-Satorras, C. Castellano, P. V. Mieghem, and A. Vespignani, \lq\lq Epidemic processes in complex networks,\rq\rq Rev. Mod. Phys. \textbf{87}, 925--979 (2015).

\bibitem{battiston2020}
F. Battiston, G. Cencetti, I. Iacopini, V. Latora, M. Lucas, A. Patania, J.-G. Young, and G. Petri, \lq\lq Networks beyond pairwise interactions: structure and dynamics,\rq\rq Phys. Rep. \textbf{874}, 1--92 (2020).

\bibitem{torres2018}
L. Torres, A. S. Blevins, D. Bassett, and T. Eliassi-Rad, \lq\lq The why, how, and when of representations for complex systems,\rq\rq SIAM Rev. \textbf{63}, 435--485 (2018).

\bibitem{watts1998}
D. J. Watts and S. H. Strogatz, \lq\lq Collective dynamics of \lq small-world\rq networks,\rq\rq Nature \textbf{393}, 440--442 (1998).

\bibitem{barabasi1999}
A.-L. Barab\'{a}si and R. Albert, \lq\lq Emergence of scaling in random networks,\rq\rq Science \textbf{286}, 509--512 (1999).

\bibitem{house2011}
T. House and M. J. Keeling, \lq\lq Insights from unifying modern approximations to infections on networks,\rq\rq
J. R. Soc. Interface \textbf{8}, 67--73 (2011).

\bibitem{zhang2013}
Y. Zhang, S. Zhou, Z. Zhang, J. Guan, and S. Zhou, \lq\lq Rumor evolution in social networks,\rq\rq Phys. Rev. E \textbf{87}, 032133 (2013).

\bibitem{javier2012}
J. Borge-Holthoefer, A. Rivero, and Y. Moreno, \lq\lq Locating privileged spreaders on an online social network,\rq\rq
Phys. Rev. E \textbf{85}, 066123 (2012).

\bibitem{tan2016}
X. Tan and K. C. Cousins, \lq\lq Herding behavior in social media networks in China,\rq\rq in \emph{Proceedings of the 22nd Americas Conference on Information Systems} (San Diego, CA, 2016).

\bibitem{bodo2016}
A. Bod\'{o}, G. Y. Katona, and P. L. Simon, \lq\lq SIS Epidemic Propagation on Hypergraphs,\rq\rq Bull. Math. Biol. \textbf{78}, 713--735 (2016).

\bibitem{suo2018}
Q. Suo, J.-L. Guo, and A.-Z. Shen, \lq\lq Information spreading dynamics in hypernetworks,\rq\rq Physica A \textbf{495}, 475--487 (2018).

\bibitem{zhang2010}
Z.-K. Zhang and C. Liu, \lq\lq A hypergraph model of social tagging networks,\rq\rq J. Stat. Mech. P10005 (2010).

\bibitem{arruda2020}
G. F. de Arruda, G. Petri, and Y. Moreno, \lq\lq Social contagion models on hypergraphs,\rq\rq Phys. Rev. Res. \textbf{2}, 023032 (2020).

\bibitem{lehmann2018}
S. Lehmann and Y.-Y. Ahn, \emph{Complex Spreading Phenomena in Social Systems} (Springer International Publishing, Switzerland, 2018)

\bibitem{schelling1971}
T. C. Schelling, \lq\lq Dynamic models of segregation,\rq\rq J. Math. Sociol. \textbf{1}, 143--186 (1971).

\bibitem{granovetter1978}
M. Granovetter, \lq\lq Threshold models of collective behavior,\rq\rq Am. J. Sociol. \textbf{83}, 1420--1443 (1978).

\bibitem{watts2002}
D. J. Watts, \lq\lq A simple model of global cascades on random networks,\rq\rq Proc. Natl. Acad. Sci. USA \textbf{99}, 5766--5771 (2002).

\bibitem{gleeson2007}
J. P. Gleeson and  D. J. Cahalane, \lq\lq Seed size strongly affects cascades on random networks,\rq\rq Phys. Rev. E \textbf{75}, 056103 (2007).

\bibitem{brummitt2012}
C. D. Brummitt, K.-M. Lee, and K.-I. Goh, \lq\lq Multiplexity-facilitated cascades in networks,\rq\rq Phys. Rev. E \textbf{85} 045102 (2012).

\bibitem{backlund2014}
V. P. Backlund, J. Saram\"{a}ki, and R. K. Pan, \lq\lq Effects of temporal correlations on cascades,\rq\rq Phys. Rev. E \textbf{89}, 062815 (2014).

\bibitem{berge1973}
C. Berge and E. Minieka, \emph{Graphs and hypergraphs} (North-Holland publishing company, Amsterdam, 1973).

\bibitem{bollobas1980}
B. Bollob\'{a}s, \lq\lq A probabilistic proof of an asymptotic formula for the number of labelled regular graphs,\rq\rq Eur. J. Comb. \textbf{1}, 311--316 (1980).

\bibitem{palla2005}
G. Palla, I. Der\'{e}nyi, I. Farkas, and T. Vicsek, \lq\lq Uncovering the overlapping community structure of complex networks in nature and society,\rq\rq Nature \textbf{435}, 814--818 (2005).

\bibitem{sorren2012}
F. Sorrentino, \lq\lq Synchronization of hyper-networks of coupled dynamical systems,\rq\rq New J. Phys. \textbf{14}, 033035 (2012).

\bibitem{wang2015}
W. Wang, M. Tang, H.-F. Zhang, and Y.-C. Lai, \lq\lq Dynamics of social contagions with memory of nonredundant information,\rq\rq Phys. Rev. E \textbf{92}, 012820 (2015).

\bibitem{huang2016}
W.-M. Huang, L.-J. Zhang, X.-J. Xu, X. Fu, \lq\lq Contagion on complex networks with persuasion,\rq\rq Sci Rep. \textbf{6}, 23766 (2016)

\bibitem{ran2020}
Y. Ran, X. Deng, X. Wang, and T. Jia, \lq\lq A generalized linear threshold model for an improved description of the spreading dynamics,\rq\rq Chaos \textbf{30}, 083127 (2020).


\end{references}
\end{document}